\documentclass[twocolumn,amsmath,amssymb,floatfix]{revtex4}

\usepackage{graphicx}
\usepackage{dcolumn}
\usepackage{bm}

\graphicspath{{Figures/}}
\setkeys{Gin}{width=\linewidth}

\begin{document}

\title{Quantum capacitance and density of states of graphene}

\author{S. Dr\"oscher} \email{susanned@phys.ethz.ch}
\author{P. Roulleau, F. Molitor, P. Studerus}
\author{C. Stampfer}
\thanks{New address: \textit{JARA-FIT and II. Institute of Physics,}}
\author{T. Ihn and K. Ensslin}
\affiliation{Solid State Physics Laboratory, ETH Zurich, 8093 Zurich, Switzerland}

\begin{abstract}

We report on measurements of the quantum capacitance in graphene as a function of charge carrier density. A resonant LC-circuit giving high
sensitivity to small capacitance changes is employed. The density of states, which is directly proportional to the quantum capacitance, is found to be
significantly larger than zero at and around the charge neutrality point. This finding is interpreted to be a result of potential fluctuations with
amplitudes of the order of 100 meV in good agreement with scanning single-electron transistor measurements on bulk graphene and transport studies on nanoribbons. 

\end{abstract}

\maketitle


The density of states of a given quantum system is important for understanding the electrical conductivity and other electronic response functions. Direct measurements of the density of states via the quantum capacitance have been successfully used, for example, on semiconductor samples containing a two-dimensional electron gas \cite{Smith1985, Stern1983} and carbon nanotubes \cite{McEuen2006}. Graphene sheets have been subject to theoretical studies of the quantum capacitance \cite{Fang2007, Zozoulenko2009} and first measurements have been carried out recently \cite{Chen2008,Xia2009}. Techniques such as scanning tunneling microscopy \cite{Yacoby2008} were applied to probe the local density of states of graphene as well. We present experiments on a structure covered with a top gate similar to those in Ref. \cite{Chen2008} and interpret our results based on the presence of disorder in the graphene device.

\begin{figure}
  \begin{center}
    \includegraphics{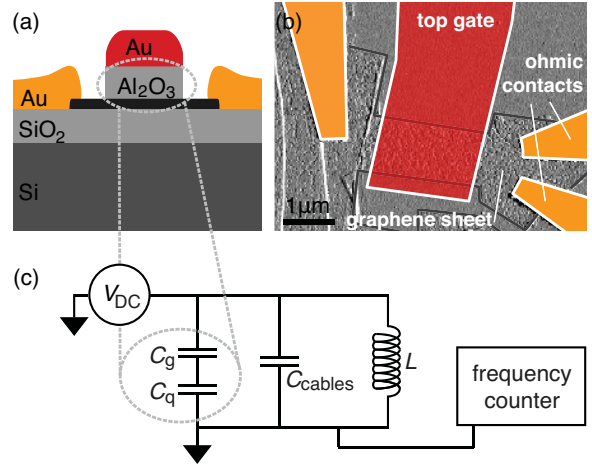}
    \caption{(a) Schematic cross section of a top gated graphene sample. The graphene flake (black) is contacted with gold electrodes (orange) and covered partly by an alumina/gold top gate (grey/red). (b) Atomic force microscope (AFM) image of the device studied here. The electrodes are colored corresponding to the scheme in (a). (c) Circuit scheme of the measured system.}
    \label{fig1}
  \end{center}
\end{figure}

The samples were fabricated using mechanical exfoliation of graphite powder and subsequent deposition onto 285 nm SiO$_2$ \cite{Novoselov2004} on an undoped silicon substrate (see Fig. \ref{fig1}(a)). It was verified that the investigated layers consisted of a single atomic sheet using optical microscopy, AFM microscopy as well as Raman spectroscopy \cite{Ferrari2006,Graf2007}. Ohmic contacts were patterned with electron beam lithography followed by metal evaporation of Cr/Au (2 nm/40 nm). In a second electron beam lithography step a mask for reactive ion etching was defined to structure the graphene sheet. The patterned topgate consisted of Ti and Au evaporated onto an Al${_2}$O${_3}$ dielectric. The alumina dielectric was obtained in cycles of depositing 1 nm of aluminum followed by a 3 min period of increased oxygen pressure in the deposition chamber leading to the complete oxidation of the thin Al-film. Repeating the cycle four times resulted in an oxide thickness of approximately 12 nm. A scheme of the cross section and an AFM image of the device investigated here are shown in Figs. \ref{fig1}(a) and (b).

All measurements were carried out in a He-bath cryostat at a base temperature of 1.7 K. Besides transport measurements through the graphene sheet using standard lock-in techniques, the capacitance between top gate and graphene sheet was recorded using an LC-circuit (see Fig. \ref{fig1}(c)). The oscillator was made of an inductor ($L$ = 100 $\mu$H) placed at room temperature and the capacitance of the wiring inside the cryostat $C_{\mathrm{cables}} \approx$ 340 pF in parallel to the sample capacitance $C_\mathrm{s}$. The sample capacitance can be decomposed into two components \cite{Smith1985, Stern1983}: the geometric capacitance $C_\mathrm{g}$ with the quantum capacitance $C_\mathrm{q}$ in series. The circuit was driven with an oscillation amplitude of 20 mV at its self-resonant frequency of 850-900 kHz. Variations of the sample capacitance were measured as a change of this frequency with a relative accuracy of 2.5$\times$10$^{-7}$, giving a sensitivity to capacitance changes of the order of 170 aF.

\begin{figure}
  \begin{center}
    \includegraphics{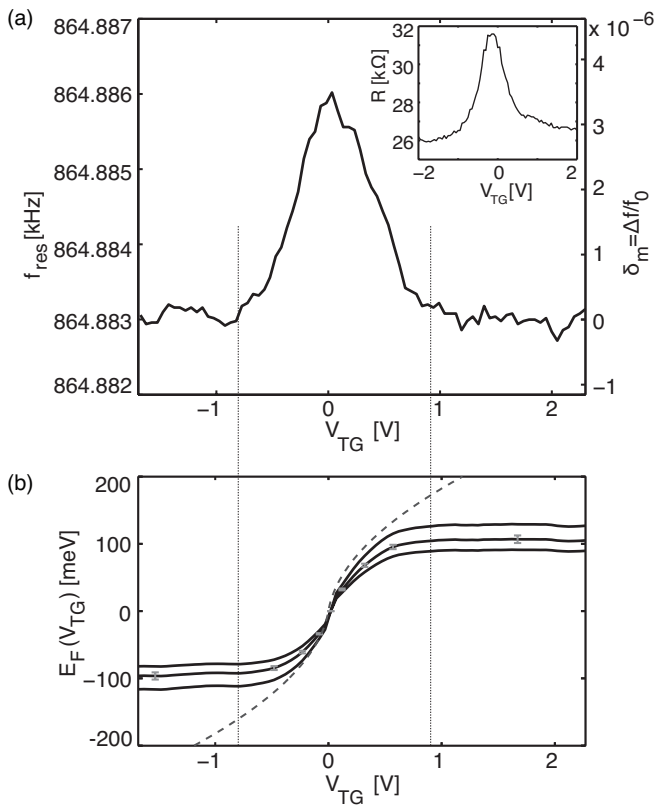}
    \caption{(a) Resonance frequency as a function of applied top gate voltage $V_\mathrm{{TG}}$. Inset: Two-point resistance obtained in transport measurements. (b) Relation between Fermi energy and top gate voltage as deduced from eq. (\ref{EF}). The different curves (black lines) are obtained for varying values of $C_\mathrm{g}$ (from outermost to innermost curve: 4.8 fF/$\mu$m$^2$, 5.8 fF/$\mu$m$^2$, 6.8 fF/$\mu$m$^2$). The dashed curve shows the theoretically expected dependence for a perfectly clean graphene sheet. To compensate for the shift of the Dirac point towards negative voltages the horizontal axis in (a) and (b) is offset by 0.3 V.}
    \label{fig2}
  \end{center}
\end{figure}

Fig. \ref{fig2} (a) shows a measurement of the resonance frequency $f_\mathrm{{res}}$ as a function of applied top gate voltage. A maximum in frequency is observed at the Dirac point where $C_\mathrm{q}$ is lowest and therefore dominates $C_\mathrm{s}$. As the density is increased $C_\mathrm{q}$ increases and its contribution to $C_\mathrm{s}$ becomes negligible beyond $C_\mathrm{q}=C_\mathrm{g}$. The constant background value of $f_\mathrm{{res}}$ far away from the charge neutrality point is hence given by $C_\mathrm{g}+C_\mathrm{{cables}}$. Since $C_\mathrm{{cables}}$ is large compared to $C_\mathrm{g}$ and the circuit inductance $L$ is known, $C_\mathrm{{cables}}$ can be determined from the measured frequency far away from the Dirac point. Considering the circuit model of our setup shown in Fig. \ref{fig1} (c) we find that the change of the capacitance $\Delta C$ compared to the background is described by  $C_\mathrm{g}^2/(C_\mathrm{g}+C_\mathrm{q})$. From the resonance condition of such an LC-circuit $\Delta C$ can be determined by expanding $f_\mathrm{0}$ for small variations leading to $\Delta C=-2(C_\mathrm{{cables}}+C_\mathrm{g})\Delta f/f_\mathrm{0}$. We define $\delta_\mathrm{m}=\Delta f(V_\mathrm{{TG}})/f_\mathrm{0}$ as the relative change in resonance frequency (see right scale in Fig. \ref{fig2} (a)). 

A two-terminal transport measurement is displayed in the inset of Fig. \ref{fig2} (a). The resistance maximum is located at the same top gate voltage as the frequency maximum in the main figure, demonstrating that these independent measurements give a consistent charge neutrality point. 


In order to transform the top gate voltage axis in Fig. \ref{fig2} (a) into Fermi energy without using assumptions about the density of states, we consider the electrostatics describing the structure. An applied top gate voltage introduces a difference in the electrochemical potentials of the top gate contact $\mu_\mathrm{M}$ and the graphene $\mu_\mathrm{G}$. The effective potential difference is given by
\begin{equation}
\mu_\mathrm{G}-\mu_\mathrm{M}= |e|V_\mathrm{{TG}}=E_\mathrm{F}+\frac{|e|^2 n_\mathrm{s}}{\epsilon \epsilon_\mathrm{0}}d +\text{const.},
\label{mu}
\end{equation}
where $E_\mathrm{F}$ is the Fermi energy, $n_\mathrm{s}$ is the charge carrier density, $d$ is the thickness of the dielectric layer and the constant term includes the work function difference of the two materials, which can be neglected for the analysis provided that it is gate voltage independent. The first term in eq. (\ref{mu}) is the chemical potential of the graphene sheet, whereas the second term describes the electrostatic potential difference between the two capacitor plates as obtained from solving Poisson's equation. The carrier density $n_\mathrm{s}$ is given by the density of states integrated from the Dirac point over all occupied energies up to $E_\mathrm{F}$. Differentiating eq. (\ref{mu}) with respect to $E_\mathrm{F}$ and identifying the geometric capacitance $C_\mathrm{g}/A=\epsilon \epsilon_\mathrm{0}/d$ and the quantum capacitance $C_\mathrm{q}(V_\mathrm{{TG}})=|e|^2\mathcal{D}(E)$ yields
\begin{equation}
\frac{\partial V_\mathrm{{TG}}}{\partial E_\mathrm{F}}= \frac{1}{|e|}\left(1+\frac{C_\mathrm{q}(V_\mathrm{{TG}})}{C_\mathrm{g}}\right).
\label{Vtg}
\end{equation}

Integrating the differential equation (\ref{Vtg}) over $V_\mathrm{{TG}}$ and including the relation between the capacitances involved and the measured frequency leads to an equation for the Fermi energy
\begin{equation}
E_\mathrm{F}(V_\mathrm{{TG}})= \frac{|e|}{\alpha}\int_{V_\mathrm{D}}^{V_\mathrm{{TG}}}\delta_\mathrm{m} (V'_\mathrm{{TG}})dV'_\mathrm{{TG}},
\label{EF}
\end{equation}
where $V_\mathrm{D}$ is the gate voltage at the Dirac point (here 0 V) and $\alpha=C_\mathrm{g}/2C_\mathrm{{cables}}$.

As the geometric capacitance can not be deduced from the data it is left as a free parameter that has to be estimated by other methods. The plate capacitor geometry of the device gives $C_\mathrm{g}\approx$ 6 fF/$\mu$m$^2$. Hall measurements on a bilayer Hall bar structure located on the same chip were used to deduce the charge carrier density and from this the geometric capacitance. A value of $C_\mathrm{g}\approx$ 5.8 fF/$\mu$m$^2$ is obtained by this additional method and hence an estimate is found for the demanded parameter.
 
Fig. \ref{fig2} (b) displays the dependence of $E_\mathrm{F}$ on $V_\mathrm{{TG}}$ for three different values of $C_\mathrm{g}$. Assuming a perfectly clean graphene sheet, one would obtain a relation indicated by the dashed curve. Close to $V_\mathrm{{TG}}$ = 0 a linear behavior is observed leading to a large increase of $E_\mathrm{F}$ for small changes of $V_\mathrm{{TG}}$ which levels off towards the boarders of the investigated bias window indicated by the dotted vertical lines. The error bars show the uncertainty of the values calculated from eq. (\ref{EF}) assuming a Gaussian probability density distribution for $\delta_\mathrm{m}$ with the width $\sigma$=2.5$\times$10$^{-7}$ determined from the variance of the noise seen in Fig. \ref{fig2} (a). With increasing $E_\mathrm{F}$ the error grows with $\sqrt{E_\mathrm{F}(V_\mathrm{{TG}}-V_\mathrm{D})}$. 

With the result in Fig. \ref{fig2} (b) at hand, allowing us to convert the $V_\mathrm{{TG}}$ axis in Fig. \ref{fig2} (a) to $E_\mathrm{F}$, we now proceed with the aim of transforming the $\Delta f/f_\mathrm{0}$ axis to a density of states. The quantum capacitance and therefore the density of states is related to the geometric capacitance and the change of the total capacitance by
\begin{equation}
C_\mathrm{q}= C_\mathrm{g}\frac{\alpha-\delta_\mathrm{m}}{\delta_\mathrm{m}}.
\end{equation}

\begin{figure}
  \begin{center}
    \includegraphics{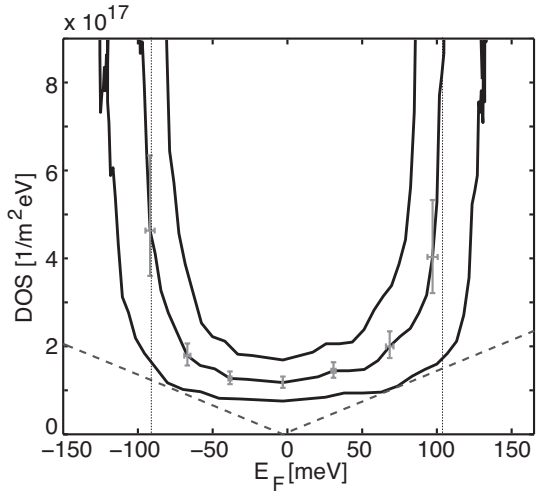}
    \caption{Density of states as a function of Fermi energy. The black solid lines show the experimental data assuming different $C_\mathrm{g}$ (from outermost to innermost curve: 4.8 fF/$\mu$m$^2$, 5.8 fF/$\mu$m$^2$, 6.8 fF/$\mu$m$^2$). Error bars are indicated and the theoretically expected density of states for a perfectly clean graphene sheet is drawn as the dashed line. The bias window defined in Fig. \ref{fig2} is plotted with dotted lines.}
    \label{fig3}
  \end{center}
\end{figure}

However, the measurement uncertainty gets very large outside the interval indicated by the vertical dashed lines in Fig. \ref{fig2} because the signal to noise ratio quickly approaches zero. A careful analysis of the data using the methods of Ref. 12 leads to the density of states $\mathcal{D}(E_\mathrm{F})$ plotted in Fig. \ref{fig3} for different $C_\mathrm{g}$ between 4.8 fF/$\mu$m$^2$ and 6.8 fF/$\mu$m$^2$. This is the main result of this paper. It shows a flat bottom in the energy interval $E_\mathrm{F}=\pm$50 meV at about 1$\times$10$^{17}$ eV$^{-1}$m$^2$ with the tendency to increase outside this interval. Error bars indicating the FWHM of the underlying probability density function for $\mathcal{D}(E_\mathrm{F})$ are shown for selected data points. The dashed line in Fig. \ref{fig3} shows the theoretical density of states of ideal graphene given by $\mathcal{D}(E_\mathrm{F})=2E_\mathrm{F}/\pi(v_\mathrm{F}\hbar)^2$ for comparison.\cite{Wallace1947} 

The data demonstrates that, in contrast to theory, there is a finite number of states close to the charge neutrality point. We attribute this observation to the presence of local potential variations within the graphene sheet on the energy scale of about 100meV.  From transport measurements on nanoribbons \cite{Stampfer2009,Molitor2009,Todd2009,Liu2009,Han2009} and studies using a scanning single-electron transistor \cite{Yacoby2008} a comparable magnitude of disorder was inferred. This suggests that the gate oxide does not induce additional disorder on this energy scale.

In conclusion, we have demonstrated a highly sensitive method for capacitance measurements using a resonant circuit. We have extracted the density of states of a single layer graphene sheet in which the density was tuned by a top gate. The magnitude of the disorder potential of the device was determined and is in good agreement with results obtained in other studies. However, in order to map features like Landau levels or the one-dimensional density of states of graphene nanoribbons \cite{Zozoulenko2009} the potential fluctuations have to be reduced drastically by improving the sample quality.

We thank K. von Klitzing, J. Martin, M. Geller and I.V. Zozoulenko for helpful discussions.

\end{document}